\documentstyle[12pt]{article}
\def\beq{\begin{equation}}
\def\eeq{\end{equation}}
\def\bea{\begin{eqnarray}}
\def\eea{\end{eqnarray}}
\def\bq{\begin{quote}}
\def\eq{\end{quote}}

\def\bq{\begin{quote}}
\def\eq{\end{quote}}

\def\bq{\begin{quote}}
\def\eq{\end{quote}}

\def\gappeq{\mathrel{\rlap {\raise.5ex\hbox{$>$}}
{\lower.5ex\hbox{$\sim$}}}}

\def\lappeq{\mathrel{\rlap{\raise.5ex\hbox{$<$}}
{\lower.5ex\hbox{$\sim$}}}}

\def\bbz{fa Z \kern-8.9pt Z}

\begin{document}

\baselineskip 24pt
\newcommand{\sheptitle}
{Large Mixing Angle MSW and Atmospheric Neutrinos from
Single Right-Handed Neutrino Dominance
and $U(1)$ Family Symmetry}

\newcommand{\shepauthor}
{S. F. King}

\newcommand{\shepaddress}
{Department of Physics and Astronomy,
University of Southampton, Southampton, SO17 1BJ, U.K.}

\newcommand{\shepabstract}
{Single right-handed neutrino dominance (SRHND) in the 23 sector
of the light effective neutrino mass matrix has been proposed as
a natural explanation for the concurrent large 23 mixing angle
and large 23 mass hierarchy.
In this paper we show how large 12 mixing angles, suitable for
the large mixing angle (LMA) MSW solution to the solar neutrino problem,
may arise from SRHND. In order to understand the
conditions for SRHND and LMA MSW we first consider
the case of one and two right-handed neutrinos,
and obtain simple analytic conditions which are then
extended to the case of three right-handed neutrinos.
We then introduce a single $U(1)$ family symmetry
and show how these analytic conditions may be translated into
$U(1)$ charge assignments and perform a systematic search
for the simplest examples.}

\begin{titlepage}
\begin{flushright}
hep-ph/9912492\\
\end{flushright}
\begin{center}
{\large{\bf \sheptitle}}
\bigskip \\ \shepauthor \\ \mbox{} \\ {\it \shepaddress} \\ \vspace{.5in}
{\bf Abstract} \bigskip \end{center} \setcounter{page}{0}
\shepabstract
\end{titlepage}

\section{Introduction}
The latest atmospheric neutrino preliminary results
from Super Kamiokande \cite{SK} are best fit by
$\nu_{\mu}\rightarrow \nu_{\tau}$ oscillations
$\sin^2 2\theta_{23} =0.99$ and
$\Delta m_{23}^2 = 3.1\times 10^{-3}\ eV^2$.
The 90\% confidence level ranges are (approximately)
$\sin^2 2\theta_{23} > 0.85$ and
$1.5\times 10^{-3}\ eV^2 <\Delta m_{23}^2 < 5.5\times 10^{-3}\ eV^2$.

Super Kamiokande is also beginning to provide important clues
concerning the correct solution to the solar neutrino problem. The
latest preliminary evidence from Super Kamiokande \cite{suzuki}
provides some support for a day-night asymmetry, as expected from
the large mixing angle MSW solution \cite{MSW}, while the energy
spectrum distortion is consistent with being flat (up to a turn up
at the end which may be accounted for by hep neutrinos) which
disfavours the small mixing angle MSW solution (SMA
MSW)\cite{MSW}, and data on the seasonal variation expected by the
vacuum oscillation (VO) approach \cite{VO} is too statistics
dominated to draw any real conclusion. Although it is too early to
draw any firm conclusions, the above indications provide some
impetus for considering the large mixing angle MSW solution (LMA
MSW) \cite{LMAMSW} to the solar neutrino problem. The best fit for
the LMA MSW solution including the day-night effect is $\sin^2
2\theta_{12} \approx 0.76$ and $\Delta m_{12}^2 \approx 2.7\times
10^{-5}\ eV^2$.

From a theoretical point of view any indication of
neutrino mass is very exciting since it represents
new physics beyond the standard model.
The see-saw mechanism \cite{seesaw} implies that the
three light neutrino masses arise from some large mass scales
corresponding to the Majorana
masses of some heavy ``right-handed neutrinos'' $N^p_R$
$M^{pq}_{RR}$ ($p,q=1,\cdots ,Z$) whose entries take values
around or below the unification scale $M_U \sim 10^{16}$ GeV.
The presence of electroweak scale Dirac mass terms $m_{LR}^{ip}$
(a $3 \times Z$ matrix) connecting the
left-handed neutrinos $\nu^i_L$ ($i=1,\ldots 3$)
to the right-handed neutrinos $N^p_R$
then results in a very light see-saw suppressed effective $3\times 3$ Majorana
mass matrix
\beq
m_{LL}=m_{LR}M_{RR}^{-1}m_{LR}^T
\label{seesaw}
\eeq
for the left-handed neutrinos $\nu_L^i$, which are the light physical
degrees of freedom observed by experiment.

If the neutrino masses arise from the see-saw mechanism then
it is natural to imagine that the Dirac neutrino mass matrices are
related somehow to those of the charged quarks and leptons,
perhaps by some relations present in unified theories,
and are thus hierarchical in nature. The existence of a physical
neutrino mass hierarchy $m_{\nu_1}\ll m_{\nu_2} \ll m_{\nu_3}$
has the attractive feature that
$\Delta m_{23}^2 \approx m_{\nu_3}^2$, and
$\Delta m_{12}^2 \approx m_{\nu_2}^2$,
which fixes $m_{\nu_3} \approx 5.6 \times 10^{-2}eV$,
and (assuming the LMA MSW solution)
$m_{\nu_2} \approx 5.2 \times 10^{-3}eV$, with rather large errors.
Thus $m_{\nu_2}/m_{\nu_3} \sim 0.1$.
In view of such a 23 mass hierarchy
the presence of a large 23 mixing angle looks a bit surprising
at first sight, especially given our experience with small
quark mixing angles.
Of course many explanations have
been put forward subsequently, including the possibility that
the 23 subdeterminant happens to be accidentally small
\cite{ellis}; that there is large mixing in the charged lepton
sector \cite{gg}; that the mixing angles are magnified
by renormalisation group effects \cite{mag}; that
the contributions to the 23 block of the light effective Majorana matrix
come predominantly from a single right-handed neutrino
\cite{SK1} which causes the 23 subdeterminant to approximately
vanish. 

The last mechanism above, called
single right-handed neutrino dominance 
(SRHND), was further developed
in the framework of theories of flavour based on a single $U(1)$
family symmetry \cite{SK2}. Within this framework general conditions
for achieving SRHND were established in terms of $U(1)$ charges
of the various fields. The analysis was directed towards
achieving the SMA MSW solution rather than the LMA MSW solution,
although the conditions for achieving either solution
were not discussed \cite{SK2}.
The purpose of the present paper is to discuss analytically
the conditions under which SRHND and LMA MSW may both be
achieved in these models, and to provide a systematic
study of the charge assignments under a single $U(1)$ family
symmetry which can satisfy both these conditions in a natural way.
In general there has been relatively
little model building dedicated to the LMA MSW solution
(compared to the SMA MSW solution) and most of the models
that are capable of accounting for both atmospheric
and LMA MSW neutrinos involve rather complicated symmetries
which stretch one's credulity. We therefore find 
the results in this paper to be quite significant,
namely that the simplest possible 
models containing a single $U(1)$ family symmetry possess
a wealth of possible charge assignments which are
capable both of describing the LMA MSW solution
and naturally able to account for the atmospheric neutrino data,
via SRHND.

We begin in the next section by introducing
$Z$ right-handed neutrinos into the standard model,
or its supersymmetric extension.
In section 3 we consider the case of a single
right-handed neutrino, developing a simple geometrical argument
which provides an intuitive understanding of the
see-saw mechanism in this case, and showing
how it can account for the atmospheric 
neutrino data. In section 4 we 
add a further subdominant right-handed neutrino as a perturbation,
and present a geometrical argument in this case, which provides
useful insight into the see-saw mechanism in this case, and 
then explore the conditions for SRHND and the LMA MSW solution.
In section 5 we discuss the case of three right-handed neutrinos,
for three different heavy Majorana textures and, using our experience
gained in the two-right-handed neutrino case, we present
conditions for LMA MSW solution.
In section 6 we introduce a single $U(1)$ family symmetry
where the $U(1)$ charges of the three right-handed neutrinos, and three
lepton doublets leads to an expansion of the
Yukawa and heavy Majorana matrix in powers of 
the Wolfenstein parameter $\lambda$ \cite{Wolf}.
In each case we obtain general conditions on the $U(1)$ charges
for achieving the
LMA MSW solution, consistent with the requirements
of SRHND and the CHOOZ constraint \cite{CHOOZ} and
tabulate the simplest examples of successful charge assignments
for each heavy texture.

\section{The (Supersymmetric) Standard Model with $Z$ Right-handed neutrinos}

The ideas in this paper apply equally well to the standard model
or one of its supersymmetric extensions involving two Higgs doublets.
To fix the notation we consider the Yukawa terms with two Higgs doublets 
augmented by $Z$ right-handed neutrinos, 
which, ignoring the quarks, are given by
\beq
{\cal L}_{yuk}=\epsilon_{ab}\left[Y^e_{ij}H_d^aL_i^bE^c_j
-Y^{\nu}_{ip}H_u^aL_i^bN^c_p + \frac{1}{2}Y_{RR}^{pq}\Sigma N^c_pN^c_q
\right] +H.c. 
\label{MSSM}
\eeq
where $\epsilon_{ab}=-\epsilon_{ba}$, $\epsilon_{12}=1$,
and the remaining notation is standard except that
the $Z$ right-handed neutrinos $N_R^p$ have been replaced by their
CP conjugates $N^c_p$ with $p,q=1,\dots, Z$
and we have introduced a singlet field $\Sigma$ whose
vacuum expectation value (VEV) induces a heavy Majorana matrix 
$M_{RR}=<\Sigma>Y_{RR}$. 
When the two Higgs doublets get their 
VEVS $<H_u^2>=v_2$, $<H_d^1>=v_1$ with $\tan \beta \equiv v_2 /v_1$
we find the terms
\footnote{In the case of the standard model we replace
one of the two Higgs doublets by the charge conjugate of the other,
$H_d=H_u^c$, and none of the results in this paper will change.}
\beq
{\cal L}_{yuk}= v_1Y^e_{ij}E_iE^c_j
+v_2Y^{\nu}_{ip}N_iN^c_p + \frac{1}{2}M_{RR}^{pq}N^c_pN^c_q +H.c.
\eeq
Replacing CP conjugate fields we can write in a matrix notation
\beq
{\cal L}_{yuk}=\bar{E}_Lv_1Y^eE_R
+\bar{N}_Lv_2Y^{\nu}N_R + \frac{1}{2}N^T_RM_{RR}N_R +H.c.
\eeq
where we have assumed that all the masses and Yukawa couplings are
real and written $Y^\ast =Y$.
The diagonal mass matrices are given by the following unitary transformations
\bea
v_1Y^e_{diag}=V_{eL}v_1Y^eV_{eR}^{\dag}={\rm diag(m_e,m_{\mu},m_{\tau})},
\nonumber \\
M_{RR}^{diag}=\Omega_{RR}M_{RR}\Omega_{RR}^{\dag}
={\rm diag(M_{R1},\ldots ,M_{RZ})},
\eea
where the unitary transformations are also orthogonal.
From Eq.\ref{seesaw} the light effective left-handed Majorana
neutrino mass matrix is
\beq
m_{LL}=v_2^2Y_{\nu}M_{RR}^{-1}Y_{\nu}^T
\label{mLL}
\eeq
Having constructed the light Majorana mass matrix it must then
be diagonalised by unitary transformations,
\beq
m_{LL}^{diag}=V_{\nu L}m_{LL}V_{\nu L}^{\dag}
={\rm diag(m_{\nu_1},m_{\nu_2},m_{\nu_3})}.
\eeq
The leptonic analogue of the CKM matrix
is the MNS matrix defined as \cite{MNS}
\beq
V_{MNS}=V_{eL}V_{\nu L}^{\dag}.
\eeq
which may be parametrised
by a sequence of three rotations about the 1,2 and 3 axes,
as in the standard CKM parametrisation,
\beq
V_{MNS}=R_{23}R_{13}R_{12}
\label{MNS}
\eeq
where
\beq
R_{23}=\left( \begin{array}{rrr}
1 & 0 & 0   \\
0 & c_{23} & s_{23}   \\
0 & -s_{23} & c_{23}
\end{array}
\right),\
R_{13}=\left( \begin{array}{rrr}
c_{13} & 0 & s_{13}   \\
0 & 1 & 0   \\
-s_{13} & 0 & c_{13}
\end{array}
\right),\
R_{12}=\left( \begin{array}{rrr}
c_{12} & s_{12} & 0   \\
-s_{12} & c_{12} & 0   \\
0 & 0 & 1
\end{array}
\right)
\label{123}
\eeq
where $s_{ij}=\sin \theta_{ij}$, $c_{ij}=\cos \theta_{ij}$, and
$\theta_{ij}$ refer to lepton mixing angles.
Note that we completely ignore CP violating phases in this paper.

\section{One right-handed neutrino}
\subsection{Geometrical argument}
We begin by recalling \cite{SK1}
our simple atmospheric neutrino model consisting of a single
right-handed neutrino $N_{R3}$ with heavy Majorana mass
$M_{RR33}=Y$ added to the (supersymmetric) standard model.
This allows Dirac couplings of the form
\beq
v_2\bar{N}_{R3}
(dN_{L1} + eN_{L2} +fN_{L3})
\label{NDirac}
\eeq
where we have written $Y^{\nu}_{13}=d$, $Y^{\nu}_{23}=e$,
$Y^{\nu}_{33}=f$.
Only one linear combination of left-handed neutrinos couples to
$\bar{N}_{R3}$ so this combination defines the
(unnormalised) mass eigenstate
\beq
\nu_{L3}= dN_{L1} + eN_{L2} +fN_{L3}
\eeq
The two linear combinations of left-handed neutrinos orthogonal
to $\nu_{L3}$ are two massless eigenstates $\nu_{L1},\nu_{L2}$.
There is clearly a massless degenerate subspace, so the
choice of basis $\nu_{L1},\nu_{L2}$ is arbitrary.
Assuming that we are in a basis where the charged leptons
are diagonal (and hence $V_{MNS}=V_{\nu_L}^{\dagger}$) the
MNS angles are given by the sequence of rotations required
to rotate the vector $\nu_{L3}$ to lie along the $N_{L3}$ axis.
This is achieved by first rotating $\nu_{L3}$
about the $N_{L1}$ axis with $\tan \theta_{23} = e/f$,
which puts the vector into the $N_{L1},N_{L3}$ plane,
\beq
R_{23}^{\dagger}
\left( \begin{array}{c}
d  \\
e   \\
f
\end{array}
\right)
=
\left( \begin{array}{c}
d  \\
0   \\
\sqrt{e^2+f^2}
\end{array}
\right)
\eeq
Then rotating the resultant vector about the
$N_{L2}$ axis with $\tan \theta_{13} = d/\sqrt{e^2+f^2}$,
which puts it along the $N_{L3}$ axis, as desired,
\beq
R_{13}^{\dagger}R_{23}^{\dagger}
\left( \begin{array}{c}
d  \\
e   \\
f
\end{array}
\right)
=
\left( \begin{array}{c}
0  \\
0   \\
\sqrt{d^2+e^2+f^2}
\end{array}
\right)
\label{R13R23}
\eeq
The final rotation $R_{12}^\dagger $ about the $N_{L3}$ axis
is clearly not uniquely defined due to the massless degeneracy.

\subsection{See-saw with $Z=1$}
The above example with a single right-handed neutrino $N_{R3}$
may be dealt with more formally as a special case
of the general see-saw model with $Z=1$, so that
$M_{RR}$ is a trivial $1 \times 1$ matrix $Y$
and $m_{LR}$ is a $3 \times 1$ column matrix
$(d, e, f)^Tv_2$.
Since $M_{RR}$ is trivially invertible
the light effective mass matrix is simply given by
\beq
m_{LL}=\frac{m_{LR}m_{LR}^T}{Y}
 =
\left( \begin{array}{ccc}
\frac{d^2}{Y} & \frac{de}{Y} & \frac{df}{Y}   \\
\frac{ed}{Y} & \frac{e^2}{Y} & \frac{ef}{Y}   \\
\frac{fd}{Y} & \frac{fe}{Y} & \frac{f^2}{Y}
\end{array}
\right)v_2^2.
\label{matrix}
\eeq
The matrix in Eq.\ref{matrix} has vanishing determinant and
rank one which implies
two zero eigenvalues. Note how the single right-handed neutrino
coupling to the 23 sector implies vanishing determinant of the 23
submatrix. This provides a natural explanation of
both large 23 mixing angles and a hierarchy
of neutrino masses in the 23 sector at the same time.
The mass matrix may be diagonalised by the same sequence of
rotations as required to put the mass eigenvector along the
$N_{L3}$ axis. Namely, the 23 rotation with $t_{23}=e/f$
clearly gives simultaneous zeroes in the 12,22,23 (and 21,32)
positions since it corresponds geometrically to
rotating $\nu_{L3}$ into the 13 plane,
\beq
R_{23}^\dagger m_{LL}R_{23}=
\left( \begin{array}{ccc}
\frac{d^2}{Y} & 0 & \frac{d\sqrt{e^2+f^2}}{Y}   \\
0 & 0 & 0   \\
\frac{d\sqrt{e^2+f^2}}{Y} & 0 & \frac{e^2+f^2}{Y}
\end{array}
\right)v_2^2.
\label{SRHN23}
\eeq
And the subsequent 13 rotation with $t_{13}=d/\sqrt{e^2+f^2}$
completes the diagonalisation since it corresponds to aligning
rotating $\nu_{L3}$ to lie along the 3 axis,
\beq
R_{13}^\dagger
R_{23}^\dagger m_{LL}R_{23}
R_{13}=
\left( \begin{array}{ccc}
0 & 0 & 0   \\
0 & 0 & 0   \\
0 & 0 & \frac{d^2+e^2+f^2}{Y}
\end{array}
\right)v_2^2.
\label{SRHN13}
\eeq
where the mass eigenvalue is
\beq
m_{\nu_3}=\frac{(d^2+e^2+f^2)}{Y}v_2^2.
\eeq

\subsection{Physical Interpretation}
According to the above arguments the mixing angles are
therefore given as,
\beq
\tan \theta_{23} = \frac{e}{f},
\ \ \tan \theta_{13} = \frac{d}{\sqrt{e^2+f^2}},
\label{angles2313}
\eeq
with $\tan \theta_{12}$ undefined.
In order to explain the atmospheric neutrino data and CHOOZ constraint
we need $\tan \theta_{23} \approx 1$ and $\tan \theta_{13} \leq \lambda$,
which implies that the three Yukawa couplings
satisfy the relations:
\beq
d\ll e \approx f
\label{1st}
\eeq
The interpretation of the atmospheric neutrino mixing is now clear.
There is a single massive neutrino whose approximately normalised form is
\beq
\nu_{L3}=  t_{13}\nu_e + s_{23}\nu_{\mu} + c_{23}\nu_{\tau}
\eeq
with a Majorana
mass $m_{\nu_3}\approx 5.6\times 10^{-2}\ eV$,
which mixes strongly with a massless state approximately given by
\beq
\nu_{L0} \approx  c_{23}\nu_{\mu} - s_{23}\nu_{\tau}
\eeq
(where we have written $N_{L1}=\nu_e$, $N_{L2}=\nu_{\mu}$,
$N_{L3}=\nu_{\tau}$, assuming we are in the diagonal charged lepton basis)
and gives rise to the approximate
two-state near maximal mixing observed by Super Kamiokande.

In order to account for the solar neutrino data a small mass perturbation
is required to lift the massless degeneracy.
In our original approach \cite{SK1,SK2}
\footnote{
Another approach \cite{DK}
which does not rely on additional right-handed neutrinos
is to use SUSY radiative corrections so that the one-loop corrected
neutrino masses are not zero but of order $10^{-5}$ eV suitable
for the vacuum oscillation solution.} we
introduced additional right-handed neutrinos in order to provide
a subdominant contribution to the effective mass matrix
in Eq.\ref{matrix}. We shall first consider the
effect of a second right-handed neutrino ${N}_{R2}$ which gives
subdominant contributions to the 23 submatrix of $m_{LL}$,
leads to a non-zero $m_{\nu_2}$, and fixes $\tan \theta_{12}$.
The conditions for achieving large mixing angles $\theta_{12}$
corresponding to the LMA MSW solution will then
become appararent. Subsequently we shall extend the analysis to the
case of three right-handed neutrinos, where our experience
gained with the two right-handed neutrinos will guide us
towards the LMA MSW solutions in that case.

\section{Two right-handed neutrinos}
\subsection{Geometrical argument}
With two right-handed neutrinos
the Dirac couplings are extended to
\beq
v_2\bar{N}_{R2}
(aN_{L1}
+ bN_{L2}
+cN_{L3})
+ v_2\bar{N}_{R3}
(dN_{L1}
+ eN_{L2}
+fN_{L3})
\label{N2Dirac}
\eeq
We shall require that the two right-handed neutrinos
have an approximately diagonal Majorana mass matrix,
since ${N}_{R3}$ must be isolated to implement the SRHND mechanism,
\beq
M_{RR}=
\left( \begin{array}{ll}
X & 0    \\
0 & Y
\end{array}
\right)
\label{RRmatrix}
\eeq
There are now two linear combinations of left-handed neutrinos which
couple to right-handed neutrinos, defined by the unnormalised states
\bea
\nu_{L2} & = & aN_{L1} + bN_{L2} +cN_{L3} \nonumber \\
\nu_{L3} & = & dN_{L1} + eN_{L2} +fN_{L3}
\eea
Neither of the states $\nu_{L2},\nu_{L3}$ are eigenvectors
since they are not mutually orthogonal.
However the vector $\nu_{L1}$ which is orthogonal to
$\nu_{L2},\nu_{L3}$ corresponds to a
massless eigenstate since it has no coupling to right-handed
neutrinos. By repeating the same rotations as in Eq.\ref{R13R23}
we can put $\nu_{L3}$ along the 3 axis.
The effect of these rotations on $\nu_{L2}$ is
\beq
R_{13}^\dagger R_{23}^\dagger
\left( \begin{array}{c}
a  \\
b   \\
c
\end{array}
\right)
=
\left( \begin{array}{c}
c_{13}a-s_{13}(s_{23}b+c_{23}c)  \\
c_{23}b-s_{23}c   \\
s_{13}a+c_{13}(s_{23}b+c_{23}c)
\end{array}
\right)
\eeq
Since $\nu_{L1}$ is orthogonal to both of $\nu_{L2},\nu_{L3}$,
we can easily construct it after the above rotations have been performed.
Up to a normalisation $\nu_{L1}$ is given in this basis by
\beq
\nu_{L1}=
\left( \begin{array}{c}
c_{23}b-s_{23}c \\
-c_{13}a+s_{13}(s_{23}b+c_{23}c)  \\
0
\end{array}
\right)
\eeq
Since $\nu_{L3}$ lies along the 3 axis, it is clear that $\nu_{L1}$
lies in the 12 plane. We may now perform a 12 rotation
to put $\nu_{L1}$ along the 1 axis. Such a rotation is given by
\beq
\tan \theta_{12}=\frac{c_{13}a-s_{13}(s_{23}b+c_{23}c)}
{c_{23}b-s_{23}c}
\label{12}
\eeq
Such a 12 rotation leaves $\nu_{L3}\sim (0,0,1)^T$ unchanged of
course, and by design gives $\nu_{L1}\sim (1,0,0)^T$.
It also rotates $\nu_{L2}$ into the 23 plane,
since $\nu_{L2}$ is orthogonal to $\nu_{L1}$,
\beq
R_{12}^\dagger R_{13}^\dagger R_{23}^\dagger
\left( \begin{array}{c}
a  \\
b   \\
c
\end{array}
\right)
=
\left( \begin{array}{c}
0 \\
s_{12}(c_{13}a-s_{13}(s_{23}b+c_{23}c)) +
c_{12}(c_{23}b-s_{23}c)   \\
s_{13}a+c_{13}(s_{23}b+c_{23}c)
\end{array}
\right)
\eeq
After having identified the massless eigenvector
$\nu_{L1}$, and rotated it to lie along the 1 axis,
we are left with the problem of finding the two massive eigenvectors
(recall that $\nu_{L2}$ and $\nu_{L3}$ are not
mutually orthogonal).
However since $\nu_{L3}$ lies along the 3 axis
and $\nu_{L2}$ is in the 23 plane we shall shortly show that
in terms of $m_{LL}$ the problem has been reduced to
diagonalising a $2\times 2$ matrix involving further 23 rotations.
Furthermore we shall show that the additional 23 rotation
that is required is actually rather small and gives only a small
correction to $\theta_{23}$ (with $\theta_{13}$ and $\theta_{12}$
unchanged).

\subsection{See-saw with $Z=2$}
In constructing $m_{LL}$
the example with two right-handed neutrinos $N_{R2},N_{R3}$
is again a special case
of the general see-saw model but now with $Z=2$.
Since $M_{RR}$ is easily invertible
the light effective mass matrix is simply obtained
\beq
m_{LL}
 =
\left( \begin{array}{lll}
\frac{d^2}{Y}+\frac{a^2}{X}
& \frac{de}{Y} +\frac{ab}{X}
& \frac{df}{Y}+\frac{ac}{X}    \\
\frac{de}{Y} +\frac{ab}{X}
& \frac{e^2}{Y} +\frac{b^2}{X}
& \frac{ef}{Y} +\frac{bc}{X}   \\
 \frac{df}{Y}+\frac{ac}{X}
& \frac{ef}{Y} +\frac{bc}{X}
& \frac{f^2}{Y} +\frac{c^2}{X}
\end{array}
\right){v_2^2}
\label{matrix2}
\eeq
We may now state the conditions for SRHND precisely
as the requirement that the $1/Y$ terms in the 23 submatrix dominate
over the $1/X$ terms in the full matrix,
\beq
\frac{e^2}{Y} \sim \frac{ef}{Y} \sim \frac{f^2}{Y} \gg \frac{xy}{X} 
\label{SRHND}
\eeq
where $x,y \in a,b,c$.
The matrix in Eq.\ref{matrix2} has vanishing determinant 
and rank two which implies
one zero eigenvalue. We now discuss the effect of the above sequence of
rotations which were followed to put $\nu_{L3}$ along the
3-axis and $\nu_{L1}$ along the 1-axis, and show that after these
rotations have been performed $m_{LL}$ is approximately diagonal
with only small additional 23 rotations required to diagonalise it.
To begin with, the 23 rotation in Eq.\ref{angles2313},
corresponding geometrically to
rotating $\nu_{L3}$ into the 13 plane, gives
\beq
R_{23}^\dagger m_{LL}R_{23}=
\left( \begin{array}{ccc}
\frac{d^2}{Y}+\frac{a^2}{X}
& \frac{c_{23}ab}{X} - \frac{s_{23}ac}{X}
& \frac{d\sqrt{e^2+f^2}}{Y}
+ \frac{c_{23}ac}{X} + \frac{s_{23}ab}{X}  \\
.
& \frac{(c_{23}b-s_{23}c)^2}{X}
& \frac{(c_{23}^2-s_{23}^2)bc}{X}
+\frac{s_{23}c_{23}(b^2-c^2)}{X}\\
.
&
.
&
\frac{e^2+f^2}{Y}
+\frac{(c_{23}c+s_{23}b)^2}{X}
\end{array}
\right){v_2^2}
\label{TRHN23}
\eeq
The reason for the cancellation of
leading $1/Y$ terms in the 12,22,23 (and 21,32) elements
of Eq.\ref{TRHN23} is the same as in Eq.\ref{SRHN23}.
The subsequent 13 rotation in Eq.\ref{angles2313},
corresponding to rotating $\nu_{L3}$ to lie along the 3-axis,
will clearly remove the  $1/Y$ terms in the 11,13 (and 31) positions,
as in Eq.\ref{SRHN13}, and will lead to a rather complicated matrix
of the form
\beq
R_{13}^\dagger R_{23}^\dagger m_{LL}
R_{23}R_{13}=
\left( \begin{array}{lll}
O(\frac{1}{X}) & O(\frac{1}{X}) & O(\frac{1}{X}) \\
. & O(\frac{1}{X}) & O(\frac{1}{X}) \\
. & . & O(\frac{1}{Y})  \\
\end{array}
\right){v_2^2}
\label{form}
\eeq
whose essential feature is that the leading $1/Y$ terms only appear
in the 33 position. The 12 rotation in Eq.\ref{12},
previously required to rotate the eigenvector
$\nu_{L1}$ to lie along the 1-axis, then gives zeroes
in the 12,13 (and 21,31) positions.
This is because the ratio of the
12/22 elements and 13/23 elements in Eq.\ref{form}
are both equal to $t_{12}$ above,
as is easy to verify. Since the eigenvalue corresponding to
$\nu_{L1}$ is massless it also leads to a zero in the 11 position.
Therefore the 12 rotation leads to a matrix of the form
\beq
R_{12}^\dagger R_{13}^\dagger R_{23}^\dagger m_{LL}
R_{23}R_{13}R_{12}=
\left( \begin{array}{lll}
0 & 0 & 0 \\
0 & O(\frac{1}{X}) & O(\frac{1}{X}) \\
0 & O(\frac{1}{X}) & O(\frac{1}{Y}) \\
\end{array}
\right){v_2^2}
\eeq
and from Eq.\ref{SRHND} we conclude that 
the remaining 23 rotations required to diagonalise the
matrix are small.

\subsection{Physical Interpretation}
According to the above arguments the mixing angles
with a dominant right-handed neutrino $N_{R3}$ and a
subdominant right-handed neutrino $N_{R2}$ are
therefore given as,
\beq
\tan \theta_{23} = \frac{e}{f},
\ \ \tan \theta_{13} = \frac{d}{\sqrt{e^2+f^2}},
\ \ \tan \theta_{12}=
\frac{c_{13}a-s_{13}(s_{23}b+c_{23}c)}{c_{23}b-s_{23}c}
\label{angles}
\eeq
The effect of the second subdominant right-handed neutrino
is to give a non-zero mass to one of the two previously massless
neutrinos (the other one remaining massless)
with an eigenvalue of order the 22
element of matrix in Eq.\ref{TRHN23},
leaving the heaviest neutrino mass unchanged at leading order,
\beq
m_{\nu_1}=0,\
m_{\nu_2} \sim \frac{(c_{23}b-s_{23}c)^2}{X}v_2^2,\
m_{\nu_3}=\frac{(d^2+e^2+f^2)}{Y}v_2^2.
\label{masses}
\eeq
Since the 12 mixing angle is the angle relevant for the MSW effect,
the conditions under which the LMA solution may be achieved corresponds to
\beq
\tan \theta_{12}
\sim \frac{a}{b-c} \sim 1
\label{2nd}
\eeq
where we have used Eq.\ref{angles} with
the approximation that $\theta_{13}$ is small and $\theta_{23}$ is large.
This condition may be traced back to the fact that the 12 angle
originates from the ratio of the 12 and 22
elements in Eq.\ref{TRHN23},
after the cancellation of the $1/Y$ terms. 
We can even state conditions for LMA MSW in terms of the
elements of the original matrix in Eq.\ref{matrix2}.
LMA MSW requires at least one of the
$1/X$ terms in the 12 or 13 elements in Eq.\ref{matrix2}
to be of order the largest of the $1/X$ terms in the
22, 23 and 33 elements of Eq.\ref{matrix2},
\beq
max\left(\frac{ab}{X},\frac{ac}{X}\right)
\sim
max\left(\frac{b^2}{X},\frac{bc}{X},\frac{c^2}{X}\right)
\label{LMAMSW}
\eeq
Being able to spot the LMA MSW case by inspection of the
original light Majorana matrix proves very useful when dealing with
more complicated situations involving three right-handed neutrinos,
as we now discuss.

\section{Three right-handed neutrinos}
With three right-handed neutrinos
the Dirac couplings are extended to
\beq
Y_{\nu}^T =
\left( \begin{array}{ccc}
a' & b' & c'\\
a & b & c\\
d & e & f
\end{array}
\right)
\label{N3Dirac}
\eeq
There are now three distinct textures for the heavy Majorana neutrino
matrix which maintain the isolation of the dominant right-handed
neutrino $N_{R3}$, namely the diagonal,
democratic and off-diagonal textures
introduced previously\cite{SK2}.
We consider each of them in turn.

\subsection{Diagonal Texture}
\beq
M_{RR}^{diag}=
\left( \begin{array}{ccc}
X' & 0 & 0    \\
0 & X & 0 \\
0 & 0 & Y
\end{array}
\right)
\label{MRRdiag}
\eeq
We may invert the heavy Majorana matrix and construct
the light Majorana matrix using the see-saw mechanism,
\beq
m_{LL}^{diag}
 =
\left( \begin{array}{ccc}
\frac{d^2}{Y}+\frac{a^2}{X}+\frac{a'^2}{X'}
& \frac{de}{Y} +\frac{ab}{X}+\frac{a'b'}{X'}
& \frac{df}{Y}+\frac{ac}{X}+\frac{a'c'}{X'}    \\
.
& \frac{e^2}{Y} +\frac{b^2}{X}+\frac{b'^2}{X'}
& \frac{ef}{Y} +\frac{bc}{X}+\frac{b'c'}{X'}   \\
.
& .
& \frac{f^2}{Y} +\frac{c^2}{X}+\frac{c'^2}{X'}
\end{array}
\right){v_2^2}
\label{matrix3}
\eeq
which compared to Eq.\ref{matrix2}
contains extra terms proportional to $1/X'$ from the
right-handed neutrino $N_{R1}$.
If these extra terms are subdominant compared to the $1/X$ terms
so that the SRHND condition in Eq.\ref{SRHND} is extended to
\beq
\frac{e^2}{Y} \sim \frac{ef}{Y} \sim \frac{f^2}{Y} \gg \frac{xy}{X} 
\gg \frac{x'y'}{X'} 
\label{SRHNDdiag1}
\eeq
where $x,y \in a,b,c$ and $x',y' \in a',b',c'$,
then the resulting neutrino spectrum will be
very similar to the case of two right-handed neutrinos with the
mixing angles as in Eq.\ref{angles} up to some
small perturbations, and a hierarchy of neutrino
masses similar to that in Eq.\ref{masses},
\beq
m_{\nu_1}\ll m_{\nu_2} \ll m_{\nu_3}.
\label{hierarchy1}
\eeq
Thus the condition for LMA MSW is as in Eq.\ref{2nd} in this case. 
On the other hand if both $1/X$ and $1/X'$ terms are
important so that the SRHND condition in Eq.\ref{SRHND} is extended to
\beq
\frac{e^2}{Y} \sim \frac{ef}{Y} \sim \frac{f^2}{Y} \gg \frac{xy}{X} 
\sim \frac{x'y'}{X'} 
\label{SRHNDdiag2}
\eeq
then the condition for LMA MSW is a simple
generalisation of Eq.\ref{LMAMSW},
\beq
max\left(\frac{ab}{X},\frac{ac}{X},\frac{a'b'}{X'},\frac{a'c'}{X'}\right)
\sim
max\left(\frac{b^2}{X},\frac{bc}{X},\frac{c^2}{X}
,\frac{b'^2}{X'},\frac{b'c'}{X'},\frac{c'^2}{X'}\right)
\label{LMAMSWdiag}
\eeq
and the lightest two neutrinos will have similar
masses rather than being hierarchical,
\beq
m_{\nu_1}\sim m_{\nu_2} \ll m_{\nu_3}.
\label{hierarchy2}
\eeq

\subsection{Democratic Texture}
\beq
M_{RR}^{dem}=
\left( \begin{array}{ccc}
X & X & 0    \\
X & X & 0 \\
0 & 0 & Y
\end{array}
\right)
\label{MRRdem}
\eeq
The democratic case (assuming the Majorana masses in the upper
block are of the same order but are not exactly equal)
will also give results qualitatively similar to the
two right-handed neutrino case in Eq.\ref{matrix2}
but the analytic expression will be more complicated,
depending on the inverse of the 2 by 2 heavy Majorana matrix.
The order of magnitude of the terms is as follows
\beq
m_{LL}^{dem}
 =
\left( \begin{array}{ccc}
\frac{d^2}{Y}+O(\frac{a^2}{X})+O(\frac{a'^2}{X})
& \frac{de}{Y} +O(\frac{ab}{X})+O(\frac{a'b'}{X})
& \frac{df}{Y}+O(\frac{ac}{X})+O(\frac{a'c'}{X})    \\
.
& \frac{e^2}{Y} +O(\frac{b^2}{X})+O(\frac{b'^2}{X})
& \frac{ef}{Y} +O(\frac{bc}{X})+O(\frac{b'c'}{X})   \\
.
& .
& \frac{f^2}{Y} +O(\frac{c^2}{X})+O(\frac{c'^2}{X})
\end{array}
\right){v_2^2}
\label{matrix4}
\eeq
In this case we might expect democracy to lead to the primed
couplings being of the same order as the unprimed couplings
\footnote{This is the case in the examples in Table 2 for instance.}
in which case the SRHND conditions are
\beq
\frac{e^2}{Y} \sim \frac{ef}{Y} \sim \frac{f^2}{Y} \gg \frac{xy}{X} 
\sim \frac{x'y'}{X} 
\label{SRHNDdem}
\eeq
where $x,y \in a,b,c$ and $x',y' \in a',b',c'$.
The mixing angles in this case have the same pattern as in
Eq.\ref{angles} but the expression for $\tan \theta_{12}$ is now
only correct qualitatively and will contain additional primed
terms. The LMA MSW condition is a simple generalisation
of that in Eq.\ref{LMAMSW}, similar to Eq.\ref{LMAMSWdiag},
\beq
max\left(\frac{ab}{X},\frac{ac}{X},\frac{a'b'}{X},\frac{a'c'}{X}\right)
\sim
max\left(\frac{b^2}{X},\frac{bc}{X},\frac{c^2}{X}
,\frac{b'^2}{X},\frac{b'c'}{X},\frac{c'^2}{X}\right)
\label{LMAMSWdem}
\eeq
The main difference compared to the two right-handed neutrino
case is that in the democratic case
if the primed terms are of order the
corresponding unprimed terms we would expect a spectrum
as in Eq.\ref{hierarchy2}.

\subsection{Off-Diagonal Texture}
\beq
M_{RR}^{off-diag}=
\left( \begin{array}{ccc}
0 & X & 0    \\
X & 0 & 0 \\
0 & 0 & Y
\end{array}
\right)
\label{MRRoff-diag}
\eeq
The off-diagonal case is qualitatively different from the
other two cases and gives
\beq
m_{LL}^{off-diag}
 =
\left( \begin{array}{ccc}
\frac{d^2}{Y}+\frac{2aa'}{X}
& \frac{de}{Y} +\frac{a'b}{X}+\frac{ab'}{X}
& \frac{df}{Y}+\frac{a'c}{X}+\frac{ac'}{X}    \\
.
& \frac{e^2}{Y} +\frac{2bb'}{X}
& \frac{ef}{Y} +\frac{b'c}{X}+\frac{bc'}{X}   \\
.
& .
& \frac{f^2}{Y} +\frac{2cc'}{X}
\end{array}
\right){v_2^2}
\label{matrix5}
\eeq
SRHND is now defined by the conditions
\beq
\frac{e^2}{Y} \sim \frac{ef}{Y} \sim \frac{f^2}{Y} \gg \frac{xx'}{X} 
\label{SRHNDoffdiag}
\eeq
where $x \in a,b,c$ and $x' \in a',b',c'$,
In this case the geometrical picture we
developed for the single right-handed
neutrino is still valid, and the mixing angles in Eq.\ref{angles2313}
remain unchanged. However the geometrical arguments
in section 4.1 are no longer applicable to this case
(where all three right-handed neutrinos play an important role) 
and the 12 rotation required in this case is obtained 
by following the proceedure in section 4.2 for the case
of two right-handed neutrinos.
The result is a more complicated expression,
\beq
t_{12}=
\frac{c_{13}[c_{23}(a'b+ab')-s_{23}(a'c+ac')]
-s_{13}[(b'c+bc')(c_{23}^2-s_{23}^2)+2s_{23}c_{23}(bb'-cc')]}
{2c_{23}^2bb'+2s_{23}^2cc'-2c_{23}s_{23}(b'c+bc')}
\label{t12comp}
\eeq
which reduces to Eq.\ref{12} in the limit $a'=a$, $b'=b$, $c'=c$.
The LMA MSW solution condition is again a straightforward
generalisation of Eq.\ref{LMAMSW},
\beq
max\left(\frac{a'b}{X},\frac{ab'}{X},\frac{a'c}{X},\frac{ac'}{X}\right)
\sim
max\left(\frac{bb'}{X},\frac{b'c}{X},\frac{bc'}{X},\frac{cc'}{X}\right)
\label{LMAMSWoffdiag}
\eeq
As in the democratic case the large 12 mixing of the right-handed
neutrinos implies a spectrum as in Eq.\ref{hierarchy2}.

\section{U(1) Family Symmetry }
\subsection{General Considerations}

Since the heavy Majorana masses and Dirac Yukawa couplings are
free parameters it is always possible to choose them to satisfy the
previous conditions. One way to constrain the theory
and give some insight into the fermion mass spectrum is via a broken
family symmetry \cite{FN}, \cite{textures}, \cite{IR}.
Such family symmetries have recently been applied to neutrino masses
\cite{ellis}, \cite{gg}, \cite{SK2}, \cite{Ramond}, \cite{Altarelli},
\cite{Barbieri}.
For definiteness we shall focus on a particular class of model based
on a single pseudo-anomalous $U(1)$ gauged family symmetry \cite{IR}.
We assume that the $U(1)$ is broken by the equal VEVs of two
singlets $\theta , \bar{\theta}$ which have vector-like
charges $\pm 1$ \cite{IR}.
The $U(1)$ breaking scale is set by $<\theta >=<\bar{\theta} >$
where the VEVs arise from a Green-Schwartz mechanism \cite{GS} 
with computable Fayet-Illiopoulos $D$-term which
determines these VEVs to be one or two orders of magnitude
below $M_U$. Additional exotic vector matter with
mass $M_V$ allows the Wolfenstein parameter \cite{Wolf}
to be generated by the ratio \cite{IR}
\beq
\frac{<\theta >}{M_V}=\frac{<\bar{\theta} >}{M_V}= \lambda \approx 0.22
\label{expansion}
\eeq

The idea is that at tree-level the $U(1)$ family symmetry
only permits third family Yukawa couplings (e.g. the top quark
Yukawa coupling). Smaller Yukawa couplings are generated effectively
from higher dimension non-renormalisable operators corresponding
to insertions of $\theta$ and $\bar{\theta}$ fields and hence
to powers of the expansion parameter in Eq.\ref{expansion},
which we have identified with the Wolfenstein parameter.
The number of powers of the expansion parameter is controlled
by the $U(1)$ charge of the particular operator.
As discussed in ref.\cite{SK2} this simple picture may
be more complicated, but it is sufficient for our purposes here.
The fields relevant to neutrino masses
$L_i$, $N^c_p$, $H_u$, $\Sigma$
are assigned $U(1)$ charges $l_i$, $n_p$, $h_u$, 
$\sigma$. From Eqs.\ref{expansion},
the neutrino Yukawa couplings and Majorana mass
terms may then be expanded in powers of the Wolfenstein parameter,
\bea
&& Y^{\nu}_{ip} \sim \lambda^{|l_i+n_p+h_u|}, \ \
M_{RR}^{pq} \sim \lambda^{|n_p + n_q + \sigma|}<\Sigma >. \ \
\label{Yukexpneut}
\eea
In dealing with the neutrino sector it is convenient to
absorb the Higgs charge $h_u$ into the
definition of the lepton charges $l_i$ so that Eq.\ref{Yukexpneut} becomes
\beq
Y^{\nu}_{ip} \sim \lambda^{|l_i+ n_p|}, \ \
M_{RR}^{pq} \sim \lambda^{|n_p + n_q + \sigma|}<\Sigma > \ \
\label{Yukexpneut'}
\eeq
The light Majorana matrix may then be constructed
from Eq.\ref{mLL}.
If we were to assume positive definite values for
$l_i+n_p$ and $n_p + n_q + \sigma$ then the modulus signs
could be dropped and the right-handed neutrino charges $n_p$
would cancel when $m_{LL}$ is constructed
using the see-saw mechanism with Eq.\ref{Yukexpneut'}  \cite{dropout}.
From the point of view of SRHND it is therefore
important that such a cancellation does not take place, and so
we shall require that at least some of the combinations
$l_i+n_p$ and $n_p + n_q + \sigma$ take negative values.
In such a case the choice of right-handed neutrino charges
will play an important role in determining $m_{LL}$,
and each particular choice of $n_p$ must be analysed
separately.

The conditions on the Yukawa couplings and heavy Majorana
masses in the SRHND approach developed earlier may now
be translated into conditions on the $U(1)$ charges,
via Eq.\ref{Yukexpneut'}. 
The heavy Majorana matrix in Eq.\ref{Yukexpneut'} is explicitly
\beq
M_{RR} \sim
\left( \begin{array}{ccc}
\lambda^{|2n_1+\sigma|} & \lambda^{|n_1+n_2+\sigma|} 
& \lambda^{|n_1+n_3+\sigma|}\\
. & \lambda^{|2n_2+\sigma|} & \lambda^{|n_2+n_3+\sigma|} \\
.  & .  & \lambda^{|2n_3+\sigma|} 
\end{array}
\right) <\Sigma >
\label{mRR}
\eeq
The conditions which ensure that the third dominant neutrino
is isolated require that the elements $\lambda^{|n_1+n_3+\sigma|}$,
$\lambda^{|n_2+n_3+\sigma|}$ be sufficiently small.
The diagonal texture condition is
\beq
|n_1+n_2+\sigma|>min(|2n_1+\sigma|,|2n_2+\sigma|), \ \ 
2|n_1+n_2+\sigma|\ge |2n_1+\sigma|+|2n_2+\sigma|)
\eeq
leading to an approximate texture as in Eq.\ref{MRRdiag}
\beq
M_{RR}^{diag} \sim
\left( \begin{array}{ccc}
\lambda^{|2n_1+\sigma|} & 0& 0\\
0 & \lambda^{|2n_2+\sigma|} & 0 \\
0  & 0  & \lambda^{|2n_3+\sigma|} 
\end{array}
\right) <\Sigma >
\label{MRRdiag2}
\eeq
The democratic texture condition is
\beq
|n_1+n_2+\sigma| = |2n_1+\sigma| = |2n_2+\sigma|
\eeq
leading to an approximate texture as in Eq.\ref{MRRdem}
\beq
M_{RR}^{dem} \sim
\left( \begin{array}{ccc}
\lambda^{|2n_1+\sigma|} & \lambda^{|n_1+n_2+\sigma|} 
& 0\\
\lambda^{|n_1+n_2+\sigma|}  & \lambda^{|2n_2+\sigma|} & 0 \\
0 & 0  & \lambda^{|2n_3+\sigma|} 
\end{array}
\right) <\Sigma >
\label{MRRdem2}
\eeq
The off-diagonal texture condition is 
\beq
|n_1+n_2+\sigma|<|2n_1+\sigma|,|2n_2+\sigma|,
\eeq
leading to an approximate texture as in Eq.\ref{MRRoff-diag}
\beq
M_{RR}^{off-diag} \sim
\left( \begin{array}{ccc}
0 & \lambda^{|n_1+n_2+\sigma|} 
& 0\\
\lambda^{|n_1+n_2+\sigma|}  & 0 & 0 \\
0 & 0  & \lambda^{|2n_3+\sigma|} 
\end{array}
\right) <\Sigma >
\label{MRRoffdiag2}
\eeq

The transpose of the Dirac Yukawa matrix 
in Eq.\ref{Yukexpneut'} is explicitly
\beq
Y_{\nu}^T \sim
\left( \begin{array}{ccc}
\lambda^{|n_1+l_1|} & \lambda^{|n_1+l_2|} 
& \lambda^{|n_1+l_3|}\\
\lambda^{|n_2+l_1|} & \lambda^{|n_2+l_2|} 
& \lambda^{|n_2+l_3|}\\
\lambda^{|n_3+l_1|} & \lambda^{|n_3+l_2|} 
& \lambda^{|n_3+l_3|}
\end{array}
\right)
\label{Ynu}
\eeq
which may be compared to the notation in Eq.\ref{N3Dirac}.
The requirement
of large 23 mixing and small 13 mixing expressed in Eq.\ref{1st}
becomes
\beq
|n_3+l_2|=|n_3+l_3|, \ \ \ \ |n_3+l_1|-|n_3+l_3|=1 \ or \ 2
\label{modulus}
\eeq

The remaining conditions for the $U(1)$ charges
depend on the specific heavy Majorana
texture under consideration. For instance the basic SRHND requirement
will be different for the three heavy textures,
as may be seen by comparing
Eqs.\ref{SRHNDdiag1}, \ref{SRHNDdiag2}, \ref{SRHNDdem}, \ref{SRHNDoffdiag}.
These conditions may be translated into conditions on the $U(1)$
charges using Eqs.\ref{MRRdiag2}, \ref{MRRdem2}, \ref{MRRoffdiag2}, \ref{Ynu},
for the various textures in Eqs.\ref{MRRdiag}, \ref{MRRdem}, \ref{MRRoff-diag},
and by requiring dominance at the order of $\lambda$ or $\lambda^2$.
Similarly the LMA MSW charge conditions
in Eqs.\ref{LMAMSWdiag}, \ref{LMAMSWdem}, \ref{LMAMSWoffdiag}
are readily translated into charge conditions.

\subsection{Examples}
We have performed a computer search over charges which satisfy all the 
conditions given above, and so provide a natural account
of the atmospheric and solar neutrinos via the LMA MSW effect.
Our first observation is that the diagonal and democratic textures
may only satisfy all the conditions if the charges are allowed
to be half-integer. This is similar to the SMA MSW results
for the two right-handed neutrino case in ref.\cite{SK2}.
Moreover all the LMA MSW solutions found have the feature that the
subdominant contributions to $m_{LL}$
are suppressed by $\lambda^2$ and also 
$\tan \theta_{13} \sim \lambda$, coresponding
physically to $m_{\nu_2}/m_{\nu_3} \sim \lambda^2$ 
and a CHOOZ angle near the current limit.
Examples of the charges in the diagonal and democratic
cases which lead to LMA MSW are given in Tables 1,2.

\begin{table}[tbp]
\hfil
\begin{tabular}{ccccccc}
\hline \hline 
             $l_1$  & $l_2$  &  $l_3$ & $n_1$ & $n_2$ & $n_3$ & $\sigma$ 
\\ \hline \hline 

             -1   &  1 & 1 & 0 & 1/2 & -1/2 & -1 \\ \hline
             -1   &  1 & 1 & 1/2 & 0 & -1/2 & -1 \\ \hline
             -1/2   &  1/2 & 1/2 & 0 & 1/2 & -1 & -1 \\ \hline
             -1/2   &  1/2 & 1/2 & 0 & 1/2 & -1/2 & -1 \\ \hline
             -1/2   &  1/2 & 1/2 & 1/2 & 0 & -1 & -1 \\ \hline
             -1/2   &  1/2 & 1/2 & 1/2 & 0 & -1/2 & -1 \\ \hline
             0   &  -1 & -1 & -1/2 & 1/2 & 1 & 1 \\ \hline
             0   &  -1 & -1 & 0 & 1/2 & 1 & 0 \\ \hline
             0   &  -1 & -1 & 0 & 1/2 & 1 & 1/2 \\ \hline
             0   &  -1 & -1 & 0 & 1/2 & 1 & 1 \\ \hline
             0   &  -1 & -1 & 1/2 & -1/2 & 1 & 1 \\ \hline
             0   &  -1 & -1 & 1/2 & 0 & 1 & 0 \\ \hline
             0   &  -1 & -1 & 1/2 & 0 & 1 & 1/2 \\ \hline
             0   &  -1 & -1 & 1/2 & 0 & 1 & 1 \\ \hline
             0   &  1 & 1 & -1/2 & 0 & -1 & -1 \\ \hline
             0   &  1 & 1 & -1/2 & 0 & -1 & -1/2 \\ \hline
             0   &  1 & 1 & -1/2 & 0 & -1 & 0 \\ \hline
             0   &  1 & 1 & -1/2 & 1/2 & -1 & -1 \\ \hline
             0   &  1 & 1 & 0 & -1/2 & -1 & -1 \\ \hline
             0   &  1 & 1 & 0 & -1/2 & -1 & -1/2 \\ \hline
             0   &  1 & 1 & 0 & -1/2 & -1 & 0 \\ \hline
             0   &  1 & 1 & 1/2 & -1/2 & -1 & -1 \\ \hline
             1/2   &  -1/2 & -1/2 & -1/2 & 0 & 1/2 & 1 \\ \hline
             1/2   &  -1/2 & -1/2 & -1/2 & 0 & 1 & 1 \\ \hline
             1/2   &  -1/2 & -1/2 & 0 & -1/2 & 1/2 & 1 \\ \hline
             1/2   &  -1/2 & -1/2 & 0 & -1/2 & 1 & 1 \\ \hline
             1   &  -1 & -1 & -1/2 & 0 & 1/2 & 1 \\ \hline \hline 
\end{tabular}
\hfil
\caption{\footnotesize
Examples of charges which satisfy the conditions of SRHND, 
and lead to a diagonal heavy Majorana texture
with hierarchical neutrino masses
$m_{\nu_2}/m_{\nu_3} \sim \lambda^2$,
large $\theta_{23}\sim 1$ (Super-Kamiokande), 
small $\theta_{13}\sim \lambda$ (CHOOZ),
and large $\theta_{12}\sim 1$ (LMA MSW).
From these charges the neutrino mass matrices of interest may be explicitly
constructed using Eqs.\ref{MRRdiag2},\ref{Ynu} together with 
Eqs.\ref{N3Dirac},\ref{MRRdiag},\ref{matrix3}.}
\end{table}

\begin{table}[tbp]
\hfil
\begin{tabular}{ccccccc}
\hline \hline 
             $l_1$  & $l_2$  &  $l_3$ & $n_1$ & $n_2$ & $n_3$ & $\sigma$ 
\\ \hline \hline 

             -1   &  0 & 0 & 1/2 & 1/2 & -1 & -1 \\ \hline
             -1   &  0 & 0 & 1/2 & 1/2 & -1/2 & -1 \\ \hline
             -1   &  0 & 0 & 1/2 & 1/2 & 0 & -1 \\ \hline
             -1   &  0 & 1 & 1/2 & 1/2 & -1/2 & -1 \\ \hline
             -1   &  1 & 0 & 1/2 & 1/2 & -1/2 & -1 \\ \hline
             -1   &  1 & 1 & 0 & 0 & -1/2 & -1 \\ \hline
             -1   &  1 & 1 & 0 & 0 & -1/2 & -1/2 \\ \hline
             -1   &  1 & 1 & 0 & 0 & -1/2 & 0 \\ \hline
             -1/2   &  1/2 & 1/2 & 0 & 0 & -1 & -1 \\ \hline
             -1/2   &  1/2 & 1/2 & 0 & 0 & -1 & -1/2 \\ \hline
             -1/2   &  1/2 & 1/2 & 0 & 0 & -1 & 0 \\ \hline
             -1/2   &  1/2 & 1/2 & 0 & 0 & -1/2 & -1 \\ \hline
             -1/2   &  1/2 & 1/2 & 0 & 0 & -1/2 & -1/2 \\ \hline
             -1/2   &  1/2 & 1/2 & 0 & 0 & -1/2 & 0 \\ \hline
             0   &  -1 & -1 & 1/2 & 1/2 & 1 & -1 \\ \hline
             0   &  -1 & -1 & 1/2 & 1/2 & 1 & -1/2 \\ \hline
             0   &  -1 & -1 & 1/2 & 1/2 & 1 & 0 \\ \hline
             0   &  -1 & -1 & 1/2 & 1/2 & 1 & 1/2 \\ \hline
             0   &  -1 & -1 & 1/2 & 1/2 & 1 & 1 \\ \hline
             0   &  1 & 1 & -1/2 & -1/2 & -1 & -1 \\ \hline
             0   &  1 & 1 & -1/2 & -1/2 & -1 & -1/2 \\ \hline
             0   &  1 & 1 & -1/2 & -1/2 & -1 & 0 \\ \hline
             0   &  1 & 1 & -1/2 & -1/2 & -1 & 1/2 \\ \hline
             0   &  1 & 1 & -1/2 & -1/2 & -1 & 1 \\ \hline
             1/2   &  -1/2 & -1/2 & 0 & 0 & 1/2 & 0 \\ \hline
             1/2   &  -1/2 & -1/2 & 0 & 0 & 1/2 & 1/2 \\ \hline
             1/2   &  -1/2 & -1/2 & 0 & 0 & 1/2 & 1 \\ \hline
             1/2   &  -1/2 & -1/2 & 0 & 0 & 1 & 0 \\ \hline
             1/2   &  -1/2 & -1/2 & 0 & 0 & 1 & 1/2 \\ \hline
             1/2   &  -1/2 & -1/2 & 0 & 0 & 1 & 1 \\ \hline
             1   &  -1 & -1 & 0 & 0 & 1/2 & 0 \\ \hline
             1   &  -1 & -1 & 0 & 0 & 1/2 & 1/2 \\ \hline
             1   &  -1 & -1 & 0 & 0 & 1/2 & 1 \\ \hline
             1   &  -1 & 0 & -1/2 & -1/2 & 1/2 & 1 \\ \hline
             1   &  0 & -1 & -1/2 & -1/2 & 1/2 & 1 \\ \hline
             1   &  0 & 0 & -1/2 & -1/2 & 0 & 1 \\ \hline \hline 
\end{tabular}
\hfil
\caption{\footnotesize
Examples of charges which satisfy the conditions of SRHND, 
and lead to a democratic heavy Majorana texture
with hierarchical neutrino masses
$m_{\nu_2}/m_{\nu_3} \sim \lambda^2$,
large $\theta_{23}\sim 1$ (Super-Kamiokande), 
small $\theta_{13}\sim \lambda$ (CHOOZ),
and large $\theta_{12}\sim 1$ (LMA MSW).
From these charges the neutrino mass matrices of interest may be explicitly
constructed using Eqs.\ref{MRRdem2},\ref{Ynu} together with 
Eqs.\ref{N3Dirac},\ref{MRRdem},\ref{matrix4}.}
\end{table}

Turning to the off-diagonal texture, we find that in this case
both integer and half-integer 
charge solutions are allowed for the 
LMA MSW solution with the SRHND interpretation of 
the atmospheric neutrino results.
The subdominant contributions to $m_{LL}$
are suppressed by $\lambda$ or $\lambda^2$,
and the CHOOZ angle may be $\tan \theta_{13}\sim \lambda$ or
$\tan \theta_{13} \sim \lambda^2$.
Examples corresponding to integer charges with
$m_{\nu_2}/m_{\nu_3} \sim \lambda^2$ 
and a CHOOZ angle $\tan \theta_{13} \sim \lambda^2$
well below the current limit are given in Table 3.

\begin{table}[tbp]
\hfil
\begin{tabular}{ccccccc}
\hline \hline 
             $l_1$  & $l_2$  &  $l_3$ & $n_1$ & $n_2$ & $n_3$ & $\sigma$ 
\\ \hline \hline 
             -2   &  0 & 0 & -2 & 1 & -1 & 1 \\ \hline
             -2   &  0 & 0 & -2 & 1 & 0 & 0 \\ \hline
             -2   &  0 & 0 & -1 & 1 & 0 & 0 \\ \hline
             -2   &  0 & 0 & 1 & -2 & -1 & 1 \\ \hline
             -2   &  0 & 0 & 1 & -2 & 0 & 0 \\ \hline
             -2   &  0 & 0 & 1 & -1 & 0 & 0 \\ \hline
             -1   &  1 & 1 & -2 & 0 & -1 & 2 \\ \hline
             -1   &  1 & 1 & 0 & -2 & -1 & 2 \\ \hline
             1   &  -1 & -1 & 0 & 2 & 1 & -2 \\ \hline
             1   &  -1 & -1 & 2 & 0 & 1 & -2 \\ \hline
             2   &  0 & 0 & -1 & 1 & 0 & 0 \\ \hline
             2   &  0 & 0 & -1 & 2 & 0 & 0 \\ \hline
             2   &  0 & 0 & -1 & 2 & 1 & -1 \\ \hline \hline 
\end{tabular}
\hfil
\caption{\footnotesize
Examples of integer charges which satisfy the conditions of SRHND, 
and lead to an off-diagonal heavy Majorana texture
with hierarchical neutrino masses
$m_{\nu_2}/m_{\nu_3} \sim \lambda^2$,
large $\theta_{23}\sim 1$ (Super-Kamiokande), 
small $\theta_{13}\sim \lambda^2$ (CHOOZ),
and large $\theta_{12}\sim 1$ (LMA MSW).
From these charges the neutrino mass matrices of interest may be explicitly
constructed using Eqs.\ref{MRRoffdiag2},\ref{Ynu} together with 
Eqs.\ref{N3Dirac},\ref{MRRoff-diag},\ref{matrix5}.
In this off-diagonal case there are many more integer solutions
with a larger $\theta_{13}\sim \lambda$ (CHOOZ),
and many more half-integer solutions not displayed.}
\end{table}

In general the dominant right-handed neutrino may be 
lighter, heavier or equal in mass to the sub-dominant right-handed
neutrinos. In order to illustrate this, and outline the general proceedure,
we construct explicitly all the neutrino mass matrices corresponding
to the first three sets of charges in Table 3.
The first set of charges in Table 3 has the dominant right-handed
neutrino being lighter by a factor of $\lambda$ than the other two, 
using Eqs.\ref{MRRoffdiag2},\ref{Ynu},
\beq
M_{RR} \sim
\left( \begin{array}{ccc}
0 & 1 & 0\\
1 & 0 & 0 \\
0 & 0  & \lambda 
\end{array}
\right) <\Sigma >, \ \ 
Y_{\nu}^T \sim
\left( \begin{array}{ccc}
\lambda^4 & \lambda^2 & \lambda^2\\
\lambda & \lambda & \lambda\\
\lambda^3 & \lambda & \lambda
\end{array}
\right)
\label{first}
\eeq
leading to dominant and sub-dominant contributions to $m_{LL}$
from $N_{R3}$ and $N_{R1},N_{R2}$ respectively using
Eqs.\ref{N3Dirac},\ref{MRRoff-diag},\ref{matrix5},
\beq
m_{LL}^{off-diag}
\sim
\left( \begin{array}{ccc}
\lambda^5 & \lambda^3 & \lambda^3    \\
\lambda^3 & \lambda & \lambda    \\
\lambda^3 & \lambda & \lambda   
\end{array}
\right)\frac{v_2^2}{<\Sigma >}
+
\left( \begin{array}{ccc}
\lambda^5 & \lambda^3 & \lambda^3    \\
\lambda^3 & \lambda^3 & \lambda^3    \\
\lambda^3 & \lambda^3 & \lambda^3   
\end{array}
\right)\frac{v_2^2}{<\Sigma >}
\eeq
In this case $N_{R3}$ gives a dominant contribution
to the 23 block from the fact that the product of the Dirac
coupling of $N_{L2},N_{L3}$ to $N_{R1}$ times the
coupling of $N_{L2},N_{L3}$ to $N_{R2}$ is
$\lambda$ times smaller than the
coupling of $N_{L2},N_{L3}$ to $N_{R3}$ squared,
with $N_{R3}$ being $\lambda$ times lighter.

The second set of charges in Table 3 has the dominant right-handed
neutrino being heavier by a factor of $\lambda$ than the other two, 
\beq
M_{RR} \sim
\left( \begin{array}{ccc}
0 & \lambda  & 0\\
\lambda  & 0 & 0 \\
0 & 0  & 1
\end{array}
\right) <\Sigma >, \ \ 
Y_{\nu}^T \sim
\left( \begin{array}{ccc}
\lambda^4 & \lambda^2 & \lambda^2\\
\lambda & \lambda & \lambda\\
\lambda^2 & 1 & 1
\end{array}
\right)
\label{second}
\eeq
leading to dominant and sub-dominant contributions to $m_{LL}$
(from $N_{R3}$ and $N_{R1},N_{R2}$ respectively)
\beq
m_{LL}^{off-diag}
\sim
\left( \begin{array}{ccc}
\lambda^4 & \lambda^2 & \lambda^2    \\
\lambda^2 & 1 & 1    \\
\lambda^2 & 1 & 1  
\end{array}
\right)\frac{v_2^2}{<\Sigma >}
+
\left( \begin{array}{ccc}
\lambda^4 & \lambda^2 & \lambda^2    \\
\lambda^2 & \lambda^2 & \lambda^2    \\
\lambda^2 & \lambda^2 & \lambda^2   
\end{array}
\right)\frac{v_2^2}{<\Sigma >}
\eeq
In this case $N_{R3}$ gives a dominant contribution
to the 23 block from the fact that the product of the Dirac
coupling of $N_{L2},N_{L3}$ to $N_{R1}$ times the
coupling of $N_{L2},N_{L3}$ to $N_{R2}$ is
$\lambda^3$ times smaller than the
coupling of $N_{L2},N_{L3}$ to $N_{R3}$ squared,
which overcomes the fact that $N_{R1}$ and $N_{R2}$ are $\lambda$ 
times lighter than $N_{R3}$.

The third set of charges in Table 3 implies that the dominant right-handed
neutrino is equal in mass to the other two, 
\footnote{This is similar to one of the examples in ref.\cite{Altarelli}.}
\beq
M_{RR} \sim
\left( \begin{array}{ccc}
0 & 1  & 0\\
1 & 0 & 0 \\
0 & 0  & 1
\end{array}
\right) <\Sigma >, \ \ 
Y_{\nu}^T \sim
\left( \begin{array}{ccc}
\lambda & \lambda & \lambda\\
\lambda^3 & \lambda & \lambda\\
\lambda^2 & 1 & 1
\end{array}
\right)
\label{third}
\eeq
leading to dominant and sub-dominant contributions to $m_{LL}$
(from $N_{R3}$ and $N_{R1},N_{R2}$ respectively)
\beq
m_{LL}^{off-diag}
\sim
\left( \begin{array}{ccc}
\lambda^4 & \lambda^2 & \lambda^2    \\
\lambda^2 & 1 & 1    \\
\lambda^2 & 1 & 1  
\end{array}
\right)\frac{v_2^2}{<\Sigma >}
+
\left( \begin{array}{ccc}
\lambda^4 & \lambda^2 & \lambda^2    \\
\lambda^2 & \lambda^2 & \lambda^2    \\
\lambda^2 & \lambda^2 & \lambda^2   
\end{array}
\right)\frac{v_2^2}{<\Sigma >}
\eeq
In this case $N_{R3}$ gives a dominant contribution
to the 23 block from the fact that the product of the Dirac
coupling of $N_{L2},N_{L3}$ to $N_{R1}$ times the
coupling of $N_{L2},N_{L3}$ to $N_{R2}$ is
$\lambda^2$ times smaller than the
coupling of $N_{L2},N_{L3}$ to $N_{R3}$ squared,
with all right-handed neutrinos being approximately degenerate.

In each of the three examples above the hierarchy of neutrino masses
$m_{\nu_2}/m_{\nu_3} \sim \lambda^2$ follows from the
fact that $N_{R3}$ dominates the contribution to the 23 block
by a factor of $\lambda^2$.
The large angle $\tan \theta_{23}\sim 1$ (Super-Kamiokande), 
and the small angle $\theta_{13}\sim \lambda^2$ (CHOOZ)
are both determined by the dominant contribution, 
while $\theta_{12}\sim 1$ (LMA MSW) arises from
the sub-dominant contribution due to the
cancellation effect of the dominant contribution
discussed earlier. Thus the order of the MNS matrix in these
examples is given from Eqs.\ref{MNS}, \ref{123} as
\beq
V_{MNS}\sim
\left( \begin{array}{ccc}
1 & 1 & \lambda^2   \\
1 & 1 & 1   \\
\lambda^2 & 1 & 1
\end{array}
\right)
\label{bimax}
\eeq
The form of $V_{MNS}$ in Eq.\ref{bimax} therefore resembles
the so-called bimaximal mixing form.
However it should be remembered that the $U(1)$ charge assignments
only give the order of all the entries in powers of the expansion
parameter $\lambda$, and so we would not expect exact bimaximal
mixing, only large 12 and 23 angles. Furthermore in obtaining our
estimates of $V_{MNS}$ we have neglected the contribution from the
charged lepton sector. Since the lepton doublets 
of the second and third families tend to have
equal $U(1)$ charges $l_2=l_3$ we would in fact expect
order unity contributions to the 23 mixing angle from the lepton
sector also. Including the charged lepton contributions we would still
expect the 23 mixing angle to remain of order unity, but
it may be enhanced or reduced depending on the phase of the
lepton contribution. All we can say is that the 23 mixing angle 
receives a contribution of order unity from both the neutrino
sector and the charged lepton sector. 
Alternatively it is possible that the contribution to 23 mixing coming from
the charged lepton sector is suppressed either because 
$l_2\neq l_3$ (for examples see near the bottom of Table 2)
or for some other dynamical reason.

Finally we note that throughout this paper we have 
labelled the dominant right-handed neutrino as $N_{R3}$,
and the subdominant right-handed neutrinos as $N_{R1},N_{R2}$.
This is completely without loss of generality since any relabelling
of the right-handed neutrinos simply corresponds to
exchanging rows of $Y_{\nu}^T$. However it is natural to 
associate rows with the largest Yukawa couplings
to the third family which justfies the labelling
of $N_{R3}$ as the dominant right-handed neutrino
in Eqs.\ref{second},\ref{third} and is consistent with 
Eq.\ref{first}. In the case of a single $U(1)$ family symmetry
all the examples of SRHND involve the natural assignment
of $N_{R3}$ as the dominant right-handed neutrino.
However in theories with two $U(1)'s$ it is possible to 
make the dominant right-handed neutrino so light that 
its Yukawa couplings are smaller than those of one of the subdominant
right-handed neutrinos, and in this case it is natural to relabel
the right-handed neutrinos $N_{R3} \leftrightarrow N_{R2}$
so that $N_{R2}$ becomes the dominant right-handed neutrino.
In such theories the Dirac mixing angles may all be small
\cite{Altarelli2}.

\section{Conclusion and Discussion}

The SRHND mechanism gives a very nice understanding of the hierarchy
$m_{\nu_2} \ll m_{\nu_3}$ due to an approximately vanishing
subdeterminant of the light mass matrix. The large 
Super Kamiokande mixing angle and small CHOOZ angle
arise from the conditions in Eq.\ref{1st}. The solar neutrino spectrum
is accounted for by including the effect of additional
right-handed neutrinos which give a small perturbation to the
effect of the dominant right-handed neutrino.
As originally formulated the SRHND mechanism was applied to the
SMA MSW solution \cite{SK1, SK2}, and it was not clear
to what extent it could be applied to the LMA MSW solution.
Although there are isolated examples of LMA MSW solutions in $U(1)$
family models in the literature \cite{Altarelli} there has been no systematic 
study of the conditions for achieving the LMA MSW solution in these models
to our knowledge, and certainly not in the framework of SRHND.
In this paper we have developed simple analytical arguments, 
beginning with the one and two right-handed neutrino cases, 
then extending the arguments to the three right-handed neutrino cases, 
which determine the conditions for SRHND and the LMA MSW solutions.
Having established the criteria for the LMA MSW
solution, we then examined whether these conditions could be
met in the framework of theories containing an abelian family symmetry.
Our main conclusion is that the conditions for SRHND and LMA MSW are very easy
to satisfy, and may be achieved in the simplest type of theory based
on a single $U(1)$ family symmetry, where the abundance
of examples is very encouraging.
From the results in Tables 1-3 all the
neutrino matrices of interest may be readily constructed,
and this was done explicitly for three examples to demonstrate that
the dominant
right-handed neutrino may be lighter, heavier or equal in mass
to the subdominant right-handed neutrinos. 

One of the basic consequences of SRHND is a hierarchical
mass spectrum with either Eq.\ref{hierarchy1} or Eq.\ref{hierarchy2}.
The fully hierarchical spectrum in Eq.\ref{hierarchy1}
requires the specific condition in 
Eq.\ref{SRHNDdiag1}, which amounts to a sort of double 
right-handed neutrino 
dominance (DRHND) which imposes an extra restriction on the theory,
so the more generic expectation is Eq.\ref{hierarchy2}.
We emphasise that there is {\em no} possibility of an inverted mass spectrum
with 
\beq
m_{\nu_1}\ll m_{\nu_2} \sim m_{\nu_3}.
\eeq
It is clear that SRHND has no preference
for the LMA MSW solution over the SMA MSW solution, 
and without going beyond the basic
framework here it is not possible to make any further predictions.
The question then naturally arises of how to include the
ideas of SRHND in a unified theory which accounts for all
the quark and lepton masses and mixing angles? Although this
question is really beyond the scope of the present paper,
we would like to make a few closing remarks on this subject.

There are two main effects which come from embedding SRHND
into a unified theory. The first is the question of RG running
of the parameters from the unification scale down to the weak scale,
and the second is that of the constraints imposed by the unified theory
on the quark and lepton Yukawa couplings. Regarding the first point,
several authors have recently analysed the effect of RG running,
and have all concluded that in the case of hierarchical
neutrino masses considered here the effects are very small
\cite{RG}. Unless the $\tau$ Yukawa coupling is quite large
at the unification scale (as may be the case for large
$\tan \beta$ ) the corrections are always quite small in general,
and the fact that our scheme here involves no fine-tuning
of any kind (as evidenced by the fact that all our arguments
have involved an order of magnitude expansion in terms
of the Wolfenstein parameter $\lambda$) means that it is
quite robust to the expected small radiative corrections
to Yukawa couplings. As far as the second point is concerned,
unification may in general give powerful restrictions on the
choice of $U(1)$ charges which may be assumed
for the leptons. For example the combination of
$SU(2)_R$ and a quark-lepton symmetry will provide a powerful
restriction on the lepton charges which must essentially
be chosen to be equal to the quark charges. However the
effect of group theoretical Clebsch coefficients, which 
must necessarily be present in order to account for other
features of the quark and lepton mass spectrum,
will be expected to have an important effect on neutrino physics.
These questions will be addressed in detail in a future 
publication \cite{KO}.

\end{document}